\documentclass[12pt,a4paper]{article}
\begin{document}
\sloppy
\title{New Developments in String Gravity and String Cosmology. A Summary 
Report}
\author{N.G. SANCHEZ\\
Observatoire de Paris, LERMA\\61, avenue de l'Observatoire\\75014 Paris, 
FRANCE}
\date{ }
\maketitle

\noindent
{\bf CONTENTS}\\ \\
\noindent
1 - Minimal String driven Cosmology and its Predictions.\\ \\ 
2 - The primordial Gravitational Wave Background in String Cosmology. \\ \\
3 - Non-Singular String-Cosmologies From Exact Conformal Field Theories. \\ \\
4 - Quantum Field Theory, String Temperature and the String Phase of De 
Sitter Space-Time. \\ \\
5 - Hawking Radiation in String Theory and the String Phase of Black Holes.\\ 
\\
6 - New Dual Relations between Quantum Field Theory and  String Regimes in 
Curved Backgrounds.\\ \\
7 - New Coherent String States and Minimal Uncertainty Principle in 
String Theory. \\ \\
\section{Minimal String Driven Cosmology and its Predictions}
In refs. \cite{is1}, \cite{is2} we constructed a minimal model for the Universe evolution fully extracted from
effective String Theory. \\ \\
By linking this model to a minimal but well established observational 
information, we proved that it gives realistic predictions on early and current energy density and its results are compatible with General Relativity. \\ \\
Interestingly enough, this model predicts the current energy density 
Omega=1 
and a lower limit Omega larger or equal 4/9. On the other hand, 
the energy density at the exit of inflationary stage is also predicted 
$\mathrm{Omega} _{\mathrm{infl}}$ = 1.\\
This result shows  agreement with General Relativity (spatially flat metric 
gives critical energy density) within an unequivalent Non-Eistenian context 
(string low energy effective equations).\\ \\
The order of magnitude of the energy density-dilaton coupled term at the 
beginning of radiation dominated stage agrees with GUT scale.\\ \\
Whithout solving the known problems about higher order corrections and 
graceful exit of inflation, we find this model closer to the observational 
Universe properties than the current available string cosmology scenarii.\\ \\
At a more fundamental level, this model is by its construction close to the 
standard cosmological evolution, and it is driven selfconsistently by the 
evolution of the string equation of state itself. \\ \\
The inflationary String Driven stage is able to reach an enough amount of 
inflation, describing a Big Bang like evolution for the metric.\\
\section{The Primordial Gravitational Wave Background in String Cosmology}
In ref.\cite{is1} we found the spectrum P(w)dw of the gravitational wave background produced in 
the early universe in string theory.\\ \\
We work in the framework of String Driven Cosmology, whose scale factors are computed with the low-energy effective string equations as well as selfconsistent solutions of General Relativity with a gas of strings as source.\\ \\
The scale factor evolution is described by an early string driven inflationary stage with an instantaneous transition to a radiation dominated stage and 
successive matter dominated stage. This is an expanding string cosmology 
always running on positive proper cosmic time.\\ \\
A careful treatment of the scale factor evolution and involved transitions is made. A full prediction of the power spectrum of gravitational waves without 
any free-parameters is given.\\ \\
We study and show explicitly the effect of the dilaton field, characteristic 
to this kind of cosmologies.\\ \\
We compute the spectrum for the same evolution description with three 
differents approachs.\\ \\
Some features of gravitational wave spectra, as peaks and asymptotic 
behaviours, are found direct consequences of the dilaton involved and not 
only of the scale factor evolution.\\ \\
\section{Non-Singular String-Cosmologies From Exact Conformal Field Theories}
In ref. \cite{vls} we constructed non-singular two and three dimensional string cosmologies  
using the exact conformal field theories corresponding to SO(2,1)/SO(1,1) and 
SO(2,2)/SO(2,1) coset models.\\ \\
All semi-classical curvature singularities are canceled in the exact theories 
for both of these cosets, but some new curvature singularities 
emerge in the quantum models.\\
However, considering different patches of the global manifolds, allows the 
construction of non-singular spacetimes with cosmological interpretation.\\ \\
In both, two and three dimensions, we constructed non-singular oscillating 
cosmologies, non-singular expanding and inflationary cosmologies including 
a de Sitter (exponential) stage with positive scalar curvature.  
Non-singular contracting and deflationary cosmologies were also constructed.\\ \\ 
We analyse these cosmologies in detail with respect to the behaviour of the 
scale factors, the scalar curvature and the string-coupling.\\ \\
The sign of the scalar curvature turns out to be changed by the quantum corrections in 
oscillating cosmologies and evolves with time in the non-oscillating cases.\\ 
\\
Similarities between the two and three dimensional cases suggest a general 
picture for higher dimensional coset cosmologies :\\
(i) Anisotropy seems to be a generic unavoidable feature,\\
(ii) cosmological singularities are generically avoided and \\
(iii) it is possible to construct non-singular cosmologies where some 
spatial dimensions are experiencing inflation while the others experience 
deflation.\\
De Sitter stage can be achieved asymptotically at early times or late 
times, but there is not a conformal coset model of this type describing a 
de Sitter background globally.
\section{Quantum Field Theory, String Temperature and the String Phase of 
De Sitter Spacetime}
The density of mass levels ${\bf \rho}$(m) and the critical temperature 
for strings in de Sitter space-time were found in ref. \cite{ms1}.\\ \\
Quantum Field Theory (QFT) and string theory in de Sitter space have been 
compared in refs. \cite{ms1} and \cite{ms3}.\\ \\
A 'Dual'-transform is introduced which relates classical to quantum string 
lengths, and more generally, QFT and string domains.\\ \\
Interestingly, the string temperature in De Sitter space turns out to be the 
Dual transform of the QFT-Hawking-Gibbons temperature.\\ \\
The quantum back reaction problem for strings in de Sitter space is addressed 
selfconsistently in the framework of the 'string analogue' model (or 
thermodynamical approach), which is well suited to combine QFT and string 
studies \cite{ms1}, \cite{ms2}, \cite{ms3}.\\ \\
We find de Sitter space-time is a self-consistent solution of the semiclassical Einstein equations in this framework. Two branches for the scalar curvature 
R$_{(\pm)}$ show up : a classical, low curvature solution (-), and a quantum 
high curvature solution (+), entirely sustained by the strings. \\ \\
There is a maximal value for the curvature R$_{\mathrm {max}}$ due to the 
string back reaction.\\ \\
Interestingly, our Dual relation manifests itself in the back reaction 
solutions : the (-) branch is a classical phase for the geometry with intrinsic temperature given by the QFT-Hawking-Gibbons temperature.\\ \\
The (+) is a stringy phase for the geometry with temperature given by the 
intrinsic string de Sitter temperature.\\ \\
2 + 1 dimensions are considered, but conclusions hold generically in D 
dimensions.\\
\section{Hawking Radiation in String Theory and the String Phase of Black 
Holes}
The quantum string emission by Black Holes is computed in the framework of the 
'string analogue model' (or thermodynamical approach), which is well suited to 
combine QFT and string theory in curved backgrounds (particulary here, as black 
holes and strings posses intrinsic thermal features and temperatures).\\ \\
The QFT-Hawking temperature T$_{\mathrm{H}}$ is upper bounded by the string 
temperature T$_{\mathrm{S}}$ in the black hole background.\\ \\
The black hole emission spectrum is an incomplete gamma function of (
T$_{\mathrm{H}}$ - T$_{\mathrm{S}}$).\\ \\
For T$_{\mathrm{H}} \ll$ T$_{\mathrm{S}}$, the spectrum yields the QFT-Hawking emission.\\ \\
For T$_{\mathrm{H}}$ near to T$_{\mathrm{S}}$, it shows highly massive string 
states dominate the emission and undergo a typical string phase transition to 
a microscopic 'minimal' black hole of mass M$_{\mathrm{min}}$ or radius 
r$_{\mathrm{min}}$ (inversely proportional to T$_{\mathrm{S}}$) and string 
temperature T$_{\mathrm{S}}$.\\ \\
The semiclassical QFT black hole (of mass M and temperature T$_{\mathrm{H}}$) 
and the string black hole (of mass M$_{\mathrm{min}}$ and temperature T$_{\mathrm{S}}$) are mapped one into another by a 'Dual' transform which links 
semi classical-QFT and quantum string regimes.\\ \\
The string back reaction effect (selfconsistent black hole solution of the 
semiclassical Einstein equations with mass M$_{+}$ (radius r$_{+}$) and temperature T$_{+}$) is computed.\\ \\
Both, the QFT and string black hole regimes are well defined and bounded:  
r$_{\mathrm{min}} \le$ r$_{+} \le$ r$_{\mathrm{S}}$, 
M$_{\mathrm{min}} \le$ M$_{+} \le$ M$_{\mathrm{S}}$, 
T$_{\mathrm{min}} \le$ T$_{+} \le$ T$_{\mathrm{S}}$.\\ \\
The string 'minimal' black hole has a life time 
$\tau _{\mathrm{min}}$ = (K/Gh) T$_{\mathrm{S}}$ $^{-3}$. \\
 \section{New Dual Relations between Quantum Field Theory and String Regimes 
in Curved Backgrounds}
We introduce a R ``Dual'' transform which relates Quantum Field Theory and 
Quantum String regimes, both in a curved background.\\ \\
This operation maps the characteristic length of one regime into the other 
and, as a consequence, maps mass domains as well.\\ \\
The Hawking-Gibbons temperature and the string maximal or critical temperature 
are dual of each other.\\ \\
If back reaction of quantum matter is included, Quantum Field and Quantum 
String phases appear, and duality relations between them manifest as well.\\ \\
This Duality is shown in two relevant examples : Black Hole and de Sitter space times, and appears to be a generic feature, analogous to the ``wave-particle'' duality.\\
\section{New Coherent String States and Minimal Uncertainty Principle in 
String Theory.}
We study the properties of {\bf exact} (all level {\it k}) quantum coherent 
states in the context of string theory on a group manifold (WZWN models). \\ \\
Coherent states of WZWN models may help to solve the unitary problem : Having 
positive norm, they consistently describe the very massive string states 
(otherwise excluded by the spin-level condition).\\ \\
These states can be constructed by (at least) two alternative procedures : 
(i) as the exponential of the creation operator on the ground state, and (ii) 
as eigenstates of the annhilation operator. In the $ k \rightarrow \infty$ 
limit, all the known properties of ordinary coherent states of Quantum 
Mechanics are recovered. \\ \\
States (i) and (ii) (which are equivalent in the context of ordinary quantum 
mechanics and string theory in flat spacetime) are not equivalent in the 
context of WZWN models.\\ \\ The set (i) was constructed by Larsen and Sanchez 
in ref. \cite{ls}. The construction of states (ii) was provided in ref.  
\cite{ls2} by the same authors. We compare the two sets and discuss their 
properties. \\ \\
We analyze the uncertainty relation, and show that states (ii) satisfy 
automatically the {\it minimal uncertainty} condition for any {\it k}; they 
are thus {\it quasiclassical}, in some sense more classical than states (i) 
which 
only satisfy it in the $ k \rightarrow \infty$ limit.\\ The modification to the Heisenberg relation is given by $2 \mathcal{H}/k$, where $\mathcal{H}$ is connected to the string 
energy.


\begin{thebibliography}{8}
\bibitem{is1} M.P. Infante and N. Sanchez, ``{\it The Primordial Gravitational Wave Background in String Cosmology}'' Phys Rev D 61, 083515 (2000).
\bibitem{is2} N. Sanchez, ``{\it String Driven Cosmology and its Predictions}''
In the 8$^{th}$ Course of the Chalonge School, NATO ASI, Kluwer Pub. vol 40, 
pp 81-102, (2001).
\bibitem{vls} H.J. de Vega, A.L. Larsen and N. Sanchez, ``{\it Non-Singular 
String Cosmologies from Exact Conformal Field Theories}'' Phys Rev D 61
066003 (2000).
\bibitem{ls} A.L. Larsen and N. Sanchez, ``{\it Quantum Coherent States in AdS and SL (2, R) WZWN Model}'' Phys Rev D 62, 046003 (2000).
\bibitem{ms1} M. Ramon Medrano and N. Sanchez, ``{\it Hawking Temperature, 
String Temperature and the String Phase of de Sitter Space-time}'' Phys Rev 
D 60, 125014 (1999).
\bibitem{ms2} M. Ramon Medrano and N. Sanchez, ``{\it Hawking Radiation in 
String Theory and the String Phase of Black Holes }'' Phys Rev D 61, 084030 
(2000).
\bibitem{ms3} M. Ramon Medrano and N. Sanchez, ``{\it New Dual Relations 
between Quantum Field Theory and String Regimes in Curved Backgrounds}'', in 
the 7$^{th}$ Course of the Chalonge School, NATO ASI, Kluwer Pub. , vol 562C, 
pp 121-130 (2001).
\bibitem{ls2} A.L. Larsen and N. Sanchez, ``{\it New Coherent String States and Minimal Uncertainty in WZWN Models}'', Nucl, Phys. B618, 301 (2001). 
\end{thebibliography}
\end{document}